# Emulating coherent backscattering in multipath tunnel systems on a near-term Quantum computer

*Shaman Bhattacharyya and Somnath Bhattacharyya[a]*

*Nano-Scale Transport Physics Laboratory, School of Physics, University of the Witwatersrand, Private Bag 3, WITS 2050, Johannesburg, South Africa*

Superconducting qubits already demonstrated potential in emulating coherent back scattering or weak localization (WL) and tunnelling phenomena however, in a real multipath system they have not been verified yet. Here we show how a double-path system can be emulated by a quantum device through construction of multiple scattering centers in closed paths (detune boxes) and tunnel barriers with a large return probability ($P_r$) of electrons. Incorporation of such arrangements of tunnel barriers can add an extra geometric phase and demonstrate Aharonov-Bohm type $\Phi_0$ and $\Phi_0/2$ oscillations in a ring and a tube, respectively. A combination of inter and intra layer tunnelling in a double-path circuit creates a phase reversal and subsequently weak anti-localization (WAL) effect with a long coherence time. Finally, angle–dependence of $P_r$ firmly establishes stability of the two-path circuit which is also associated with a phase reversal due to the inter-path resonance.

Several years before a quantum simulator has been used in the group of John Martinis (Chen *et al.*, Nat. Commun. 2014) to emulate coherent back scattering popularly known as weak localization (WL) phenomena for one path system [1]. Also an opposite trend namely weak anti-localization (WAL) with an electronically created phase reversal was claimed which inspired simulation of a larger chiral circuit [1,2]. This phenomenon is commonly observed in weakly disordered materials consists of grains separated by tunnel barriers which effectively form a multipath system. WL effect can generally be verified from negative magnetoresistance (MR) behavior however, multilayer effect can change the trend significantly depending on the interlayer coupling [3-9]. The interlayer scattering and tunnelling is responsible for the transport which could have multiple origins like strong spin-orbit coupling, Kondo scattering and a resonance between paths yielding a phase reversal [10]. Although the effect of WL process has been applied in fitting transport data the time and angle dependence of quantum interference in a multipath anisotropic system need a clear demonstration. The tunnelling process is explained through superposition of states which can effectively lead to the suppression of the WL effect by means of the interband transitions [11] which can improve the understanding of transport in complex microstructures [12-17].

A direct way to observe WL is Aharonov-Bohm (AB) effect in a confined ring-like structure where the phase shift $\phi_s$ relates to the magnetic field ($\varphi$) and flux $\Phi$ as $\phi_s = \varphi = 2\pi\Phi/\Phi_0$ which produces oscillations with the period of $\Phi_0$ and also of $\Phi_0/2$ in a tube [4]. The phase shift being the sum of the product of detuning frequency ($\delta$) and time ($\tau$) accounts the disorder of the system through the product of wavevector and mean free path and also the angle between these two vectors. The problem of negative MR with an angle dependent oscillation in disordered superlattices with permitted in-plane anisotropy, motivated several transport experiments is treated theoretically on the ground of extension to the multiband regime [3] and the in-plane anisotropy [18-20]. Here we show periodic oscillations in MR by inserting an AB phase in the circuit created by qubits. The connection between the two paths can be established through (i) direct coupling using Toffoli gates (Fig. 1a-c) (ii) tunnel barrier using one CNOT gate (Fig. 2) and (iii) a pair of tunnel barriers enclosing space creating an AB phase (Fig. 3). Finally, from angle ($\phi$ & $\theta$) dependence for all cases we show the anisotropy in a two path WL process (Fig. 4). We aim to develop the possible structure of a topologically protected system through demonstration of WAL effect.

Model (Two path WL): We simulate multilayer structures (Fig. 1d-g) on IBM QE by creating circuits to emulate WL, having incorporated with inter and intrapath tunnelling. Our circuit consists of 4-7 qubits that undergo a detuning process. They are connected to the register qubits, also to the input qubit (Q1) symmetrically (Fig. 1c). WL process is effectively described by the Tavis-Cummings (T-C) Hamiltonian [1,2]:
$$H = \sum_{r=1}^{2}(\omega_r a^\dagger a) + \sum_{i=1}^{5}(\hbar\omega_i \sigma_i^+ \sigma_i^-) + \sum_{i=1}^{5} \hbar g(a^\dagger \sigma_i^- + a\sigma_i^+), \quad (1)$$
Here $r \in [1;2]$ and $i \in [1;5]$ represent the register qubits and phase qubits (including Q1) with the frequencies $\omega_r$ and $\omega_i$, respectively. With the annihilation operator for the resonator $a$ and the qubit spin operator $\sigma_i^{-,+}$ and the coupling strength of the qubit-resonator $g$ the T-C model describes a two dimensional system (Fig. 1b). If such system is represented in a 3D system then the qubits where the WL process occurs only rotate around the z-axis described by $[\sigma^+, \sigma^-] = \sigma_z$. For $\phi$ and $\theta$ dependence qubits were rotated along x and y axis, respectively.

Tunnelling: Tunnelling between two dots is given as
$$H_T = \sum_{\alpha\beta\sigma} t_{\alpha\beta} \, e^{iV_{mn}t} a_{\alpha\sigma}^+ a_{\beta\sigma} + h.c., \quad (2)$$
$V_{mn}$ is the applied voltage to the barrier ($\alpha$, $\beta$ being the single particle states) [12]. The coordinate $y$ runs along the interface $S$. The transverse coordinates $z$ in the adjacent grains are defined in such a way that at the interface $z = z' = 0$. The tunnel amplitude is expressed with the eigenfunctions ($X_\alpha$, $X_\beta$) of an electron within dots:
$$t_{\alpha\beta} = a \int dy d_z X_\alpha^*(y,z) d_{z'} X_\beta(y,z')|_{z=z'=0}, \quad (3).$$
This is the equivalent to a system with 2 potential wells separated with a barrier [21-23]. The potential barrier is treated via the Suzuki-Trotter decomposition in an efficient way of simulating tunnelling between 2 qubits [22]. Unlike a CPHASE gate used in a spin system the tunnelling terms can be added to T-C model describing the WL through a CNOT gate. In our previous work we showed that NV centers could successfully use this technique [23]. A double barrier system is constructed from two bi-level systems where spin can tunnel between two superposed Bell states [21]. Through the rotation of qubits superposition of four states can be constructed. A similar flip of $y$ component of momentum has been described in inter-granular tunnelling [12,14]. Whilst oscillatory tunnelling features are supressed at one junction two more tunnel junctions can enclose an area (Fig. 2b, inset) to produce a geometric phase and AB like oscillations as shown by Gobulev *et al.* (Fig. 3a,b) [14].

---

[a] *Electronic mail: somnath.bhattacharyya@wits.co.za*

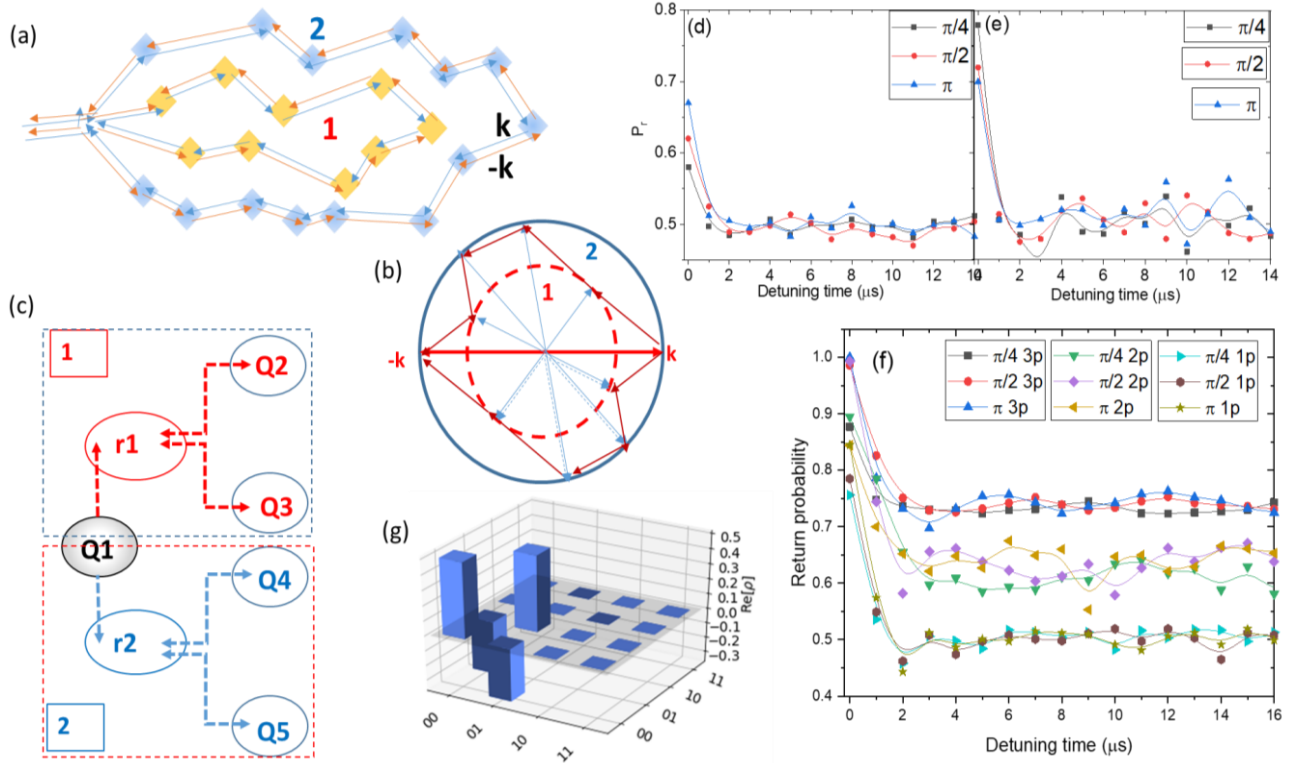

*Fig 1 (a) Double path WL model shows multiple backscattering (b) (-k to k) in paths 1 (red) & 2 (blue) with a resonance-like interband transition. (c) Qubits arrangements Q1-Q5 and r1, r2 on IBM QE. Classical simulation of WL in (d) one and (e) two path systems, the later shows an improved return probability measured at different angles ($\pi/4$-$\pi$). (f) Emulation of WL in one, two and three path (p) system shows an increase of the return probability with paths. (g) Quantum state tomography showing the Real part of density of |00> states for 1 path WL with high $P_R$.*

*AB ring and tube*: Two chaotic quantum dots separated by spaces (the lowest energy parameters) and connected by two tunnelling junctions (Fig. 3a, inset) create a ring-shaped gap in the middle of the dots [14]. Now a magnetic flux ($\Phi$) is induced and this flows through the ring shaped structure allowing passing electrons acquire different phases ($\phi_s$) which should be measured from experiments [2]. AB effect also involves in creating a hole in an insulator through a Hikami box containing advanced and retarded transportation modes and tunnelling happens through a band insulator between 2 metals [13]. An $h/2e$ oscillation results due to this setup. Instead of using a magnetic field ($\varphi$) we consider a mechanism consisting of a CNOT gate and a detuning box placed at the control of this gate generates the pulses required to stabilise the AB ring which is generated at the target of the CNOT gate. WL takes place in this system as electrons propagate across the insulator and is also associated with a back propagation described by the advanced and retarded Green's function [3-11]. We replace this propagation in our system with the forward and backscattering of qubits through detuning. Therefore we construct a system comprising of 2 path WL (T-C model) with tunnelling within each path. These 2 paths are separated by an AB ring which is stabilised by pulses fired from a detuning box. The paths are joined together at one end and the resistance correction $\delta R(\Phi)$ = -$R_{AB}$ $cos(4\pi\Phi/\Phi_0)$–$R_{WL1}$–$R_{WL2}$ shows both AB oscillations and WL effects for WL path 1 and 2 [14] (Fig. 3 a,b). For tube the periodicity changes to $\Phi_0/2$ i.e. half of the AB phase for a ring (Fig. 3 c,d).

## Results and Discussion

*WL:* All of these circuits were implemented on the IBM Melbourne device (see Methodology). We first show the simulated return probability ($P_R$) of the ground state *vs.* the total detuning time (*t*) equivalent to the resistance of the system and the magnetic field, respectively [1]. For both WL the temperature of the system remains unchanged. For both 1 & 2 paths and accumulated data the $P_R$ reaches its maximum when *t* = 0 $\mu s$ where the time reversal symmetry (TRS) remains unaffected by pulses. As the total pulse duration moves away from zero the $P_R$ decreases exponentially until it reaches a value of approximately 0.75 and 0.90 (emulated) for one and two paths (Fig. 1 f) instead of simulated values 0.62 and 0.75 in Fig. 1 d, e, respectively. The peak to 0.99 for the 3 path system which is much larger than the maximum $P_r$ value obtained 0.6 in the work of Chen *et al.* [1]. Then the $P_R$ fluctuates randomly with a slow decline. As the number of paths were increased to 2 or 3 $P_R$ increased in general (Fig. 1f). For one path $P_r$ decays steeper manner than 2 or 3 paths as shown in multilayer structures [9] and also in single and bi-layer graphene [11]. This indicates that there is constructive interference between the two paths of the circuit. These results are consistent with Bergman and others who showed time dependent WL in a multipath system [3-9]. The variation of $P_R$ with angles ($\phi$) $\pi/4$-$\pi$ for 1 path is found to be opposite to 2 path WL due to resonance related phase shift between two paths (see also in Fig. 4). The estimated coherence time is ~$\mu s$ obtained from $\phi_s$ the operating frequency ($\delta$~0.2 GHz) and separation (~1 $\mu m$) between qubits. The high density of the |00> state at the origin for WL is seen in Fig. 1g.

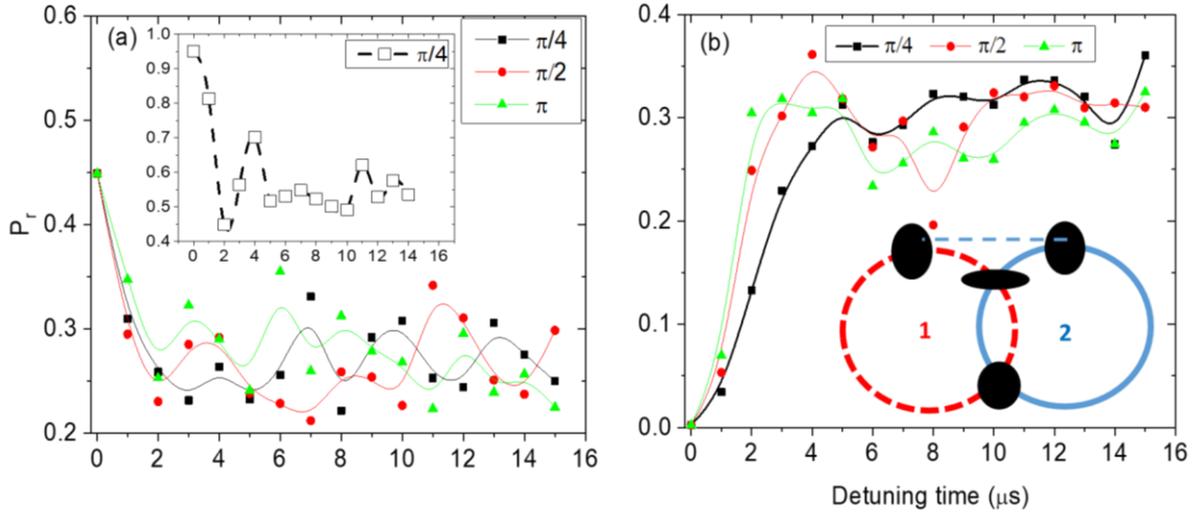

*Fig 2 (a) Interpath and (b) Combined inter and intra path tunnelling measured at different angles shows strong WAL effect compared to purely intrapath tunnelling (inset a) showing random fluctuations and supressed WL effect. (inset b) Tunnelling between two paths shown as black dots.*

*Tunnelling:* We demonstrate the intrapath in one path and interpath tunnelling between two paths of the circuit using one and two qubits, respectively in Fig. 2a and inset. With the tunnelling induced in the interlayer system the $P_R$ drops rapidly from its maximum probability at $t = 0$ until it reaches a very low value of 0.3 followed by a strong fluctuation in a random fashion (Fig. 2 a,b inset). Comparing $P_R$ measured at different angles ($\phi = \pi/4 - \pi$) a difference not only by the peak at $t = 0$ but also at the fluctuation point was found which are slightly different but random. This picture does not improve much for intralayer tunnelling (Fig. 2a, inset) which indicates high level of noise in this tunnelling process. Overall the work shows the effect of tunnelling on WL process as suggested by Blanter and others [12]. However we extend the work to a closed (triangular) path created by inter and intra-path tunnel junctions (Fig. 2b, inset) which accumulates an extra ($\pi$) phase and shows WAL effect (Fig. 2b). The $P_R$ of the ground state *vs.* $t$ for both paths shows WAL effect as the $P_R$ reaches its minimum at $t = 0$ with a value of ~0.0016. It is followed by an exponential increase in time until it reaches a stable value of around 0.3 and then fluctuates randomly with a slow increase which is exactly opposite to the WL. The frequency was kept constant for this experiment. For an inspection of the minor details of our circuit we consider the same basic details and only differences caused by changing a single parameter ($\phi$). We see that for different values of $\phi$ the smaller the $\phi$-value the rate of decay is lower as compared with the other 2 angles where the rate of decay is much higher. Despite all of this all three values of $\phi$ tend towards the same value where they fluctuate randomly (not a periodic oscillation). This increase in the $P_R$ also shows the effects of noise in the circuit. Like WL, WAL gives an indication of decoherence with the lower probabilities showing the least decoherence. These results are very different from Chen *et al.* where a WAL like effect was created electronically through a $\pi$ phase change in the circuit [1] rather introducing a geometric phase.

*AB ring and tube:* For the simulation of the AB ring with the two path WL we measure the $P_R$ for the ground state *vs.* $t$ that was used to generate and sustain the AB ring (in $\mu s$) at different values of $\phi$ (Fig. 3 a,b). These oscillations are a result of the $\pi/4$ pulses which continuously drive the qubit in a ($\Phi_0$) periodic manner (Fig. 3b, inset). Since these oscillations happen in conjunction with the exponential decay of the $P_R$ of the ground state the amplitude of the oscillations also decrease until they completely disappear. For the accumulated data as well as the two paths we notice again that the $P_R$ reaches its maximum at $t = 0$. Then as in normal WL $P_R$ decays until it reaches a stable value around 0.5 as the time is increased. However, unlike WL when the $P_R$ reaches its minimum $t$ does not fluctuate randomly, instead it oscillates (with a period of $\Phi_0$) as a Bessel function which does not reach close to the original peak at all. Using tunnel barriers a tube consists of two rings is constructed (Fig. 3c, inset) which shows oscillations with a period of $\Phi_0/2$ (Fig. 3c,d). A remarkable similarity can be found with the work of Sharvin and Altshuler *et al.* and others with the present work [15-18]. We then focus on the small details of the region away from the central peak. While the overall concept of the circuit remains the same we notice that as we increase the value of $\phi$ the time taken for the oscillations to cease along the fluctuating point decreases. In the work of *Roushan et al.* a synthetic magnetic field was generated by circulating photons around three coupled qubits arranged a triangular loop which formed a part of a larger qubit lattice [2]. This method of generating the magnetic field is very similar to the formation of an AB ring due to the presence of a closed loop. In our work due to the arrangement of the qubits in the computer we are forced to use a square shaped loop however, this yields similar results to the triangular loop proposed by *Roushan et al.* [2]. This can be shown by the presence of periodic ($h/2e$) oscillations and the estimated effective area (S) is obtained from $\Delta\varphi_s \cdot \mathbf{S} = h/2e$ as ~$\mu s^2$. By changing the ring into a cylinder consists of two loops connected by tunnel junctions the period of oscillations can be $\Phi_0/2$ (Fig. 3 c,d). Such an effect was described as flux-dependent part of WL correction originating from the phase coherence between multiple scattering paths which can be obtained in the multi-qubit arrangements. By maintaining the TRS in the paths the relative phase becomes $4\pi\Phi/\Phi_0$. [15-17,24]. Due to noise these oscillations decrease with $t$.

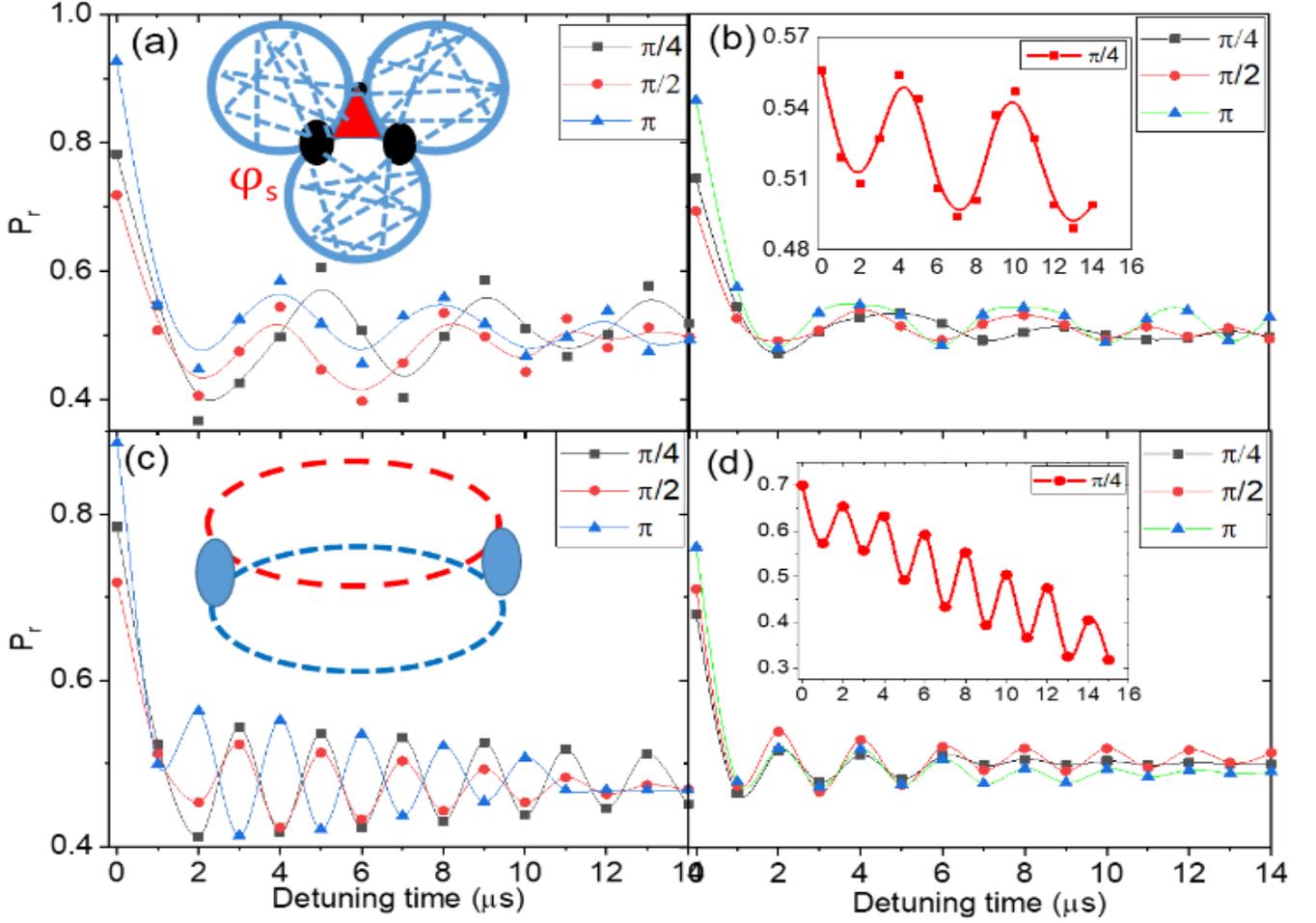

*Fig 3(a)* Emulation and *(b)* Simulation of periodic oscillations ($\Phi_0$) and a supressed WL effect for an AB ring. *Inset a*: Multipath WL model with tunnelling applied to a multi-quantum dot system forms an AB phase $\varphi_s$. *Inset b*: Simulation of AB oscillations without considering WL. *Fig. (c)* Emulation and *(d)* Simulation of $\Phi_0/2$ oscillations for a tube. *Inset c*: Multipath tunnel junction forms an AB tube. *Inset d*: Simulating oscillations without WL for a tube.

**Angle dependence:** The magnetic field is effectively rotated by qubit rotation around the x and y-axis (Inset of Fig. 4a, 4d) which shows periodic changes in $\phi$ and $\theta$ (Fig. 4 a-d). For one path the amplitude shows wide variations compared to the two path with a $\pi/2$ phase shift arising from the resonance between 2 paths (Fig. 4 a,c). The AB ring also creates a phase shift in the two path WL circuit (Fig. 4 b,d).

Destruction of interference occurs between these processes resulting in the similar variation of angle dependence in one path circuit however, with a much lower amplitude. We first examine the $\phi$-dependence and notice that for one path, tunnelling and the AB ring it follows a *cosine* pattern with different $P_R$ amplitudes. However, for two path WL the dependence is shifted by $\pi/2$ to the right representing a *sine* function. This *cosine* pattern is caused by the fact that with increase of $\phi$ we increase the trace distance away from the ground state. We measure the qubit from the |0⟩ state. As the distance increases the lower $P_R$ becomes as the state tends towards |1⟩. The difference in angle dependence between one and two path WL is caused by the fact that when we interfere the two registers onto the readout qubit they excite the readout qubit but then due to the second excited resonator the qubit collapses back to the ground state. In one path WL the readout qubit does not collapse back to the ground state and the difference between these states cause the different angle dependences between two path and one path WL. Focusing on the $\theta$-dependence of the above mentioned phenomena we notice that dependency of the phenomena is exactly shifted by $\pi/2$. Now we rotated the $\theta$ angle by $\pi$ at the beginning of circuit which means that we rotate $\theta$ from the excited state where $P_R$ is 0.00. This results in increase of $P_R$ with $\theta$. This is different from $\phi$ dependence where we do not rotate $\phi$ by $\pi$ and therefore we see that $P_R$ is at its maximum for $\phi = 0$.

Angle dependent change of phase can not only supports this model but also produces additional information of a complex material most importantly time dependence of the interference process which was not experimentally shown otherwise. The result for the anisotropic single-band case with the replacement of the Boltzmann constant by its multiband counterpart can be verified from the angle dependent MR [19,20,25]. On the ground of extension to the multiband regime and the in-plane anisotropy of the analytical approach developed before by Bryksin and Kleinert within a self-consistent theory of Anderson localization [25]. It is shown that the extension prevents from the problem reduction to a single relaxation time scaling MR behavior since a set of intra- and inter-band relaxation times emerges, affecting components of both WL corrections to the MR and the Boltzmann diffusion tensor itself. We show the angle dependent MR which differentiates single to double bands through the additional phase shift due to interband scattering (Fig. 4 a,c). The space created by two tunnel junctions is compared with the AB rings in Fig. 4b,d. The aim of the present work is to develop the modification of this formalism which involves both multiband effect and in-layer anisotropy of the superlattice as expected, a two-parameter scaling description of corresponding corrections to MR [26] which has also been described before [20]. We show the oscillations in MR here through emulation which finds a remarkable similarity with the experimental and theoretical description of polar and azimuthal angle dependence in a multilayered system [20]. Overall the experimentally observed anisotropic MR can be described by a combination of $\phi$ and $\theta$ dependence [20]. This work may find applications in optics such as tilt angle dependence of the modulated interference effects [27].

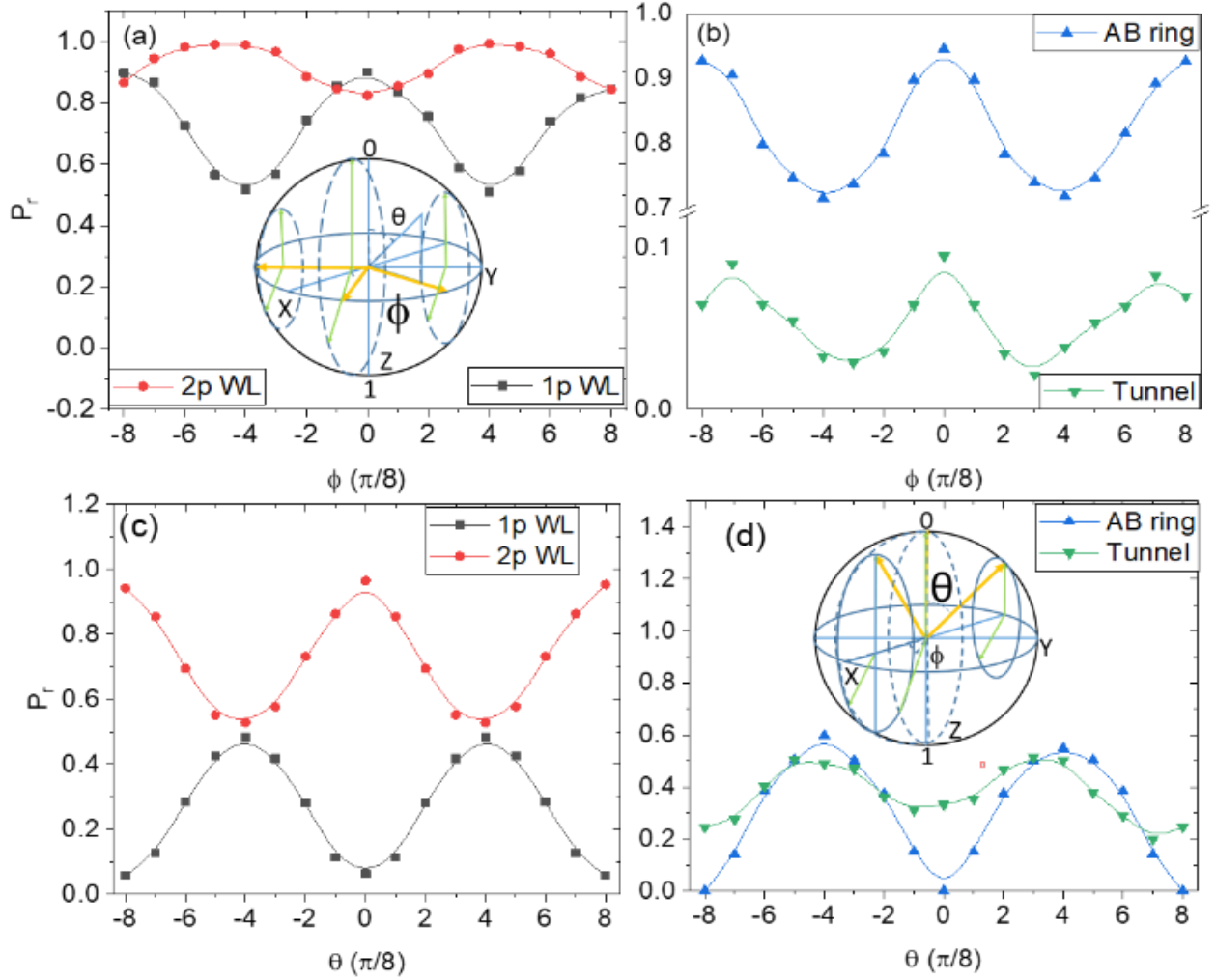

*Fig 4 Angle ($\theta$ and $\phi$) dependent variation of $P_r$ at the initial time for **(a,c)** 1 path WL, 2 path WL, **(b,d)** AB ring and tunnelling. 2 path WL (with a phase shift) and tunnelling have less variation than 1 path WL and AB ring. **Inset:** Block sphere showing **(a)** $\phi$ and **(b)** $\theta$ angle dependence.*

Topological effects: Now there is a topological approach to emulate WAL. The detuning box is similar to that proposed by *Chen et al.* as both of these boxes drive the qubits around the z-axis [1]. The external circuit is crucial when we consider the system in terms of condensed matter physics. The DRCNOT gate is an interesting gate due to the fact it can be implemented on both topological and non-topological quantum systems which makes it an ideal gate for the simulation of both WL and WAL. Previous approaches involve either only topological systems (ASI also known as the chiral spin-flip gate [28] gates) or purely non-topological systems (iSWAP and Simultaneous iSWAP [1,2]). In addition to the WL and conductance fluctuation [24] one can use this approach to describe other topological phenomena such as spin-orbit interactions and the Kondo effect [29,30] by implementing the DRCNOT gate on a Kagome lattice.

In summary, in this Letter we have explained the origin and nature of tunnelling in a multipath system which can be applied to a topological insulating system [29,30]. Starting from WL effect associated with random scattering the return probability was improved by 15% in double paths. We observe a topological phase through incorporation of a geometric structure and WAL effect by overcoming the effect of disorder. While addition of intralayer tunnel junction reduces the return probability in the circuit it can be improved through interlayer tunnelling and yields AB like oscillations. An additional phase was introduced through the interaction between paths in the double path WL circuit which is clearly seen from the angle dependence. In future the tunnelling term between layers can be constructed based on a spin impurity center with a spin flipping feature which undergoes a double operation to account spin-orbit coupling. However, our quantum simulation is able to show the validity of condensed matter models proposed by a large number of communities towards supressed disorder effect through developing a topological phase.

*Acknowledgements*:

We acknowledge use of IBM Q for this work. The views expressed are those of the authors and do not reflect the official policy or position of IBM or the IBM Q team. SB is thankful to J. Martinis, Yu. Chen and P. Roushan for introducing the subject and showing the experiment. SB would like to thank Y. Hardy and D. Churochkin for sharing ideas and stimulating discussions.